# Moderating Effects of Retail Operations and Hard-Sell Sales Techniques on Salesperson's Interpersonal Skills and Customer Repurchase Intention

Prathamesh Muzumdar

College of Business and Information Systems

Dakota State University

Madison, South Dakota, United States

Ganga Prasad Basyal

College of Business and Information Systems

Dakota State University

Madison, South Dakota, United States

Piyush Vyas

College of Business and Information Systems

Dakota State University

Madison, South Dakota, United States

**Abstract**

Salesperson's interpersonal skills have always played an important role in influencing various stages of customer's purchase decision. With the increase in retail outlets and merchandisers, retail operations have taken a pivotal role in influencing the salesperson's sales practices and customer's purchase decisions. This study tries to examine the influence of retail operations and hard-selling strategies on the relationship between salesperson's interpersonal skills and customer repurchase intention. Salesperson's interpersonal skills are the trained and tacit competencies that a salesperson employs to improve customer relationship and sales performance. Many organizations prefer skill training as a strategic approach to increase sales. Though successful in its objective, salesperson's skill training fails to attract repetitive purchases due to unavoidable extraneous factors. It has become a necessity to understand the role of extraneous factors like retail operations and hard-selling sales strategies on sales performance. This study examines the negative influence of retail operations on the relationship between salesperson's interpersonal skills and customer repurchase intention. The findings suggest that retail operations significantly moderate the relationship between salesperson's interpersonal skills and customer repurchase intention. We also find that hard-sell sales techniques play a significant moderating role in negatively influencing customer repurchase intention. This study has important implications for retailers and sales managers.

## 1. Introduction

Frontline employees play an important role in building a relationship with customers. Their abilities to understand the need and interpret those needs play an important part in building long term relationships. The important aspect of two-way communication is listening which helps to build long-term relationships (Ingram et al., 1992). Effective listening is influenced by various factors like customer orientation, adaptive selling, satisfaction, trust, salesperson performance, and intention to future interaction (Goad, 2014). Customer orientation (CO) is an approach that helps the employees to satisfy customer's long-term needs (Saxe & Weitz, 1982). Customer orientation is a key component of market orientation and plays an important role in a firm's performance (Kirca et al., 2005). Customer orientation plays an integral part in developing the interpersonal skills of an employee (Pelham, 2009). In general, customer orientation influences salesperson's interpersonal skills in a very positive way.

Customer repurchase intention is considered as an important factor to influence the customers' attitude towards product repurchase (Dick & Basu, 1994). Attitudinal and behavioral loyalty are two measures of loyalty (Pan et al., 2012; Slack et al., 2020). Behavioral loyalty is largely affected by the interpersonal skills of a salesperson (Naumann et al., 2000; Handayani et al., 2020)). Listening remains to be an integral part of a salesperson's interpersonal skills (Ramsey & Sohi, 1997). This study tries to examine the effects of salesperson's interpersonal skills on customer repurchase, as customer orientation improves a salesperson's listening ability, it indirectly influences customer loyalty in the form of product repurchase.

Retail operations and sales strategies are two extraneous factors that indirectly influence sales performance. Service management theory plays an important role in developing service strategies in the retail industry (Ho & Chung, 2020). Service strategies further lead to developing strategic retail operations resulting in better sales performance (Ilyas et al., 2020). There exists a relatable effect in this chain of interdependencies which further can lead to damaging the sales performance. Hard-selling strategies have been common for the last five decades. Once considered as a strong indicator of sales performance, today it has lost the lack-luster when creating the emotional appeal to the customer. These two extraneous factors if not attained on the strategy level can lead to failed sales performance.

This research aims to address the existing gaps in research regarding the specific impact of a salesperson's interpersonal skills on customer repurchase intention. It empirically examines a research framework based on consumer psychology and sales theories. This study provides a new research framework for a salesperson's interpersonal skills theory, taking into account the environmental and operational factors of the retail industry. It further provides a theoretical perspective of the research and eventually uses multivariate statistics to empirically solve the research questions. The research further concludes taking into account the industrial implications of the research framework.

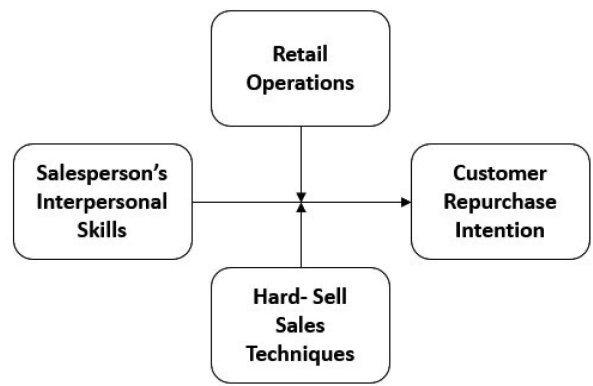

Figure 1. Theoretical Framework

## 2. Literature Review and Theoretical Framework

Salesperson's interpersonal skills theory indicates that better relationships with customers can be built on performance through proactive behaviors that directly influence sales performance and purchase outcomes (Johnson et al., 2003). Proactive reactions and behaviors in sales are described as making adaptations to the surrounding environment, responding quickly to requests, gathering queries, and responding to customers (Eckert, 2006; Johnson et al., 2003). Proactive reactions at work involve learning from experiences and training that is instituted in the organization (Goad, 2014). Salesperson's interpersonal skills are mainly categorized into four categories active listening, persuasion, delegation, and stewardship.

Salesperson listening has been the most important factor in building long-term relationships with retail customers. Any purchase decision is always affected by the relationship that a customer develops with the frontline employee and this dyadic relationship is based on the listening abilities of the salesperson (Ramsey & Sohi, 1997). Most personalized experiences for customers come through relational communications further leading to repurchases (Javed & Wu, 2020). Relationships are built by giving attention to the customer's needs throughout the sales process. Sales personal are evaluated based on their listening skills.

Salesperson persuasion is responsible for leading persuasive sales. The sole purpose of persuasive sales is to increase the probability that the customer will engage in favorable behavior at some point in time (Mowen, 1998). Persuasion selling efforts play an important role when the customer fails to take the initiative to close the sale (Du et al., 2020). The purchase decision making process is influenced by persuasion. Funkhouser (1984) presented a practical theory of persuasion where he pointed out that various persuasion techniques can be employed to influence the purchase decision-making ability of the buyer at different stages of decision making. Persuasion in selling plays an important eventually pivoting the whole decision-making process (Simanjuntak et al., 2020). This research takes into consideration the salespersons' persuasion skills as a part of interpersonal skills.

Delegation in personal selling and sales management mainly refers to point of price delegation. How salespeople handle the purchase? And the role price discounting plays in it are two important factors for price delegation (Bello et al., 2020). In price delegation, a salesperson can offer a discount to a customer based on his previous knowledge of the market (Curina et al., 2020). It is argued that local salespeople have a better understanding of the local markets and can set better purchase prices to negotiate with customers (Anderson et al., 2008). Due to the closeness to the market salespeople are in better informed about the local conditions and better understand customer's willingness to pay the price.

Customer stewardship control (CSC) is a marketing concept in which salespeople take moral responsibility and ownership of customers' overall welfare (Singh & Tiwari, 2019). Salespeople stewardship is a practice that provides responsive customer service in the best interest of the customer (Tandon et al., 2020). Traditional control theories are based on the fundamental of sales and profit-making, where salespeople are seen as rational, opportunistic actors who complete sales in the best interest of their organizations rather than the customers' (Eisenhardt, 1989). Stewardship theory contradicts this belief by emphasizing customer welfare and organizational structure (Hernandez, 2008). Salespeople stewardship plays an important role in governing and guiding customer purchase decisions.

Retail operations are an important aspect of retail revenue. They are the environmental factors that influence a customer's tacit experience (Gabler et al., 2019). Many purchase decisions are shaped by how customers perceive the situation? And what role retail operations play in it? The most important factor in retail operations is efficiency and highly efficient operations are time-constrained. Time is an important factor, where salespeople's astute reply to customer needs can positively influence the purchase decision.

Hard-sell sales technique also known as Hard-sell strategies came into existence at the onset of depression when supply outpaced demand (Liu et al., 2020). As the economic crisis was widespread consumer disposable income was limited in comparison to excess income, hence consumers were more careful with their expenditures (Fullerton, 1988). Hard-sell sales technique approaches are direct and information-based, focused more on price, benefits, and sales promotions (Butt et al., 2017). Hard-sell sales techniques are defined by many aspects of organizational mechanisms to increase sales and salesperson's performance (Ben Amor, 2019). In this study, we focus on three aspects of hard-selling techniques, pushy salespeople, overwhelming information, and lack of emotional appeal. We have explored previous work done by scholars where they have considered these three factors as the most influencing factors impeding hard-selling strategy.

The research study adds to the current literature of salespeople's interpersonal skills and customer repurchase by introducing retail operations as an environmental factor and hard-sell sales technique as a sales strategy. Retail operations differ by the magnitude of their product offerings and request processing capabilities. It has come to our notice that very few studies have taken into consideration the role retail operations play in influencing the customer's purchase decision. Most literature has focused on retail operations from the point of view of operations research and supply chain management. On the other hand, hard-sell sales techniques are seen from the point of predictors of failed sales strategy influencing sales performance, rather than a moderator. As per our understanding to date, no research has accounted for retail operations and hard-sell sales techniques as influencing factors driving customer's purchase decision. This study adds to the current literature on salesperson's interpersonal skills by adding retail operations and hard-sell sales techniques as essential influencers to narrate customer's repurchase decision.

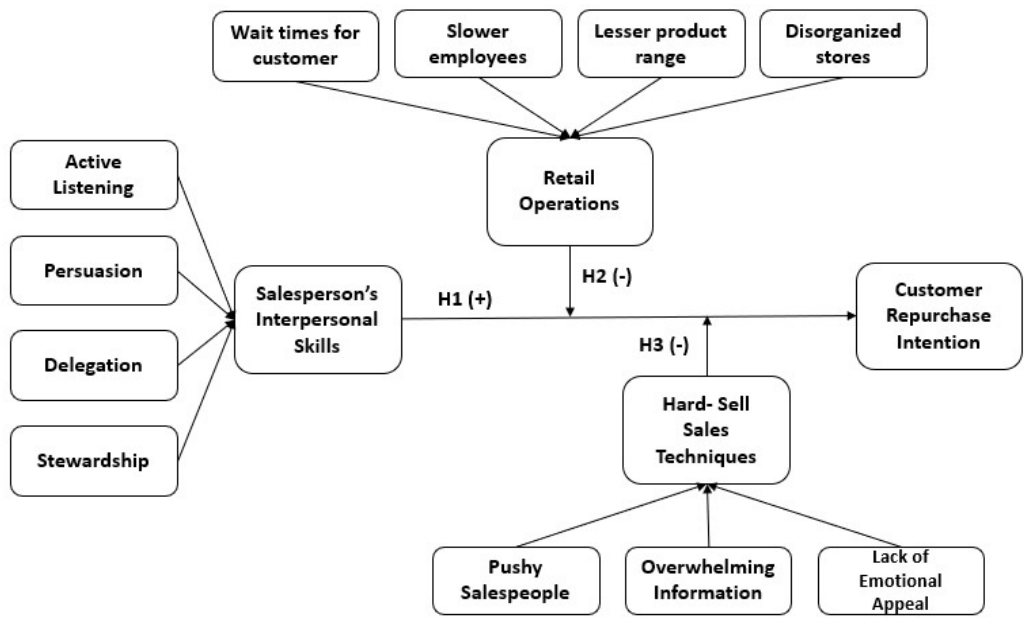

Figure 2. Research Framework

## 3. Research Question and Hypothesis

Previous researchers indicate that there is a relationship between customer repurchase intention and salesperson's interpersonal skills (Babakus et al., 1999). Better interpersonal skills result in better service eventually resulting in a satisfied customer (Bhuian et al., 2005). A salesperson's positive influence towards the customer will result in customer's frequent visits to the retailer further resulting in repurchases (Boles et al., 1997). Customer repurchase is an important factor for customer retention and sales performance (Udayana et al., 2019). Frontline employees contribute most to customer retention as they interact with customers creating the first and the last impression as the primary point of contact.

Current literature on customer repurchase intention and salesperson's interpersonal skills have failed to explore the direct relationship between them when it comes to understanding it from the point of retail operations and hard-sell sales techniques in the retail industry. We in our research extend this theory by exploring the role of retail operations and hard-sell sales techniques on the relationship between salesperson's interpersonal skills and customer repurchase intentions. We are motivated to propose the following research questions in this study.

**RQ1:** Do salesperson's interpersonal skills influence customer repurchase intention?

**RQ2:** Do retail operations moderate the influence of salesperson's interpersonal skills on customer repurchase intention?

**RQ3:** Do hard-sell sales techniques moderate the influence of a salesperson's interpersonal skills on customer repurchase intention?

*3.1 Customer Repurchase Intention*

Customer repurchase intention is measured when a customer desire to repurchase a product or products from the same retailer. It's a behavioral and attitudinal decision taken to address a person's need. Customer repurchase intention is the locus of customer retention and can be affected by multiple intrinsic and extrinsic factors, importantly, it is a sensitive pivot across which the fulcrum of loss and profit dwindles. This study's focal point is in understanding how customer repurchase intention can be influenced by the interpersonal skills of the Frontline sales employees (Ingram et al., 1992). Throughout this study, the influence of other variables will be measured over customer repurchase intention. Customer repurchase intention is the outcome variable.

*3.2 Salesperson's Interpersonal Skills*

**H1:** Salesperson's interpersonal skills positively influence customer repurchase intention.

Paying attention to customers is an important aspect of making a customer satisfied with your service. Customers are willing to purchase more if the salesperson is able to attend to their needs and provide them the service they look forward to (Saxe & Weitz, 1982). Salesperson's ability to listen to customer needs and comprehend that information to develop a sense of customer's purchase intention can help the salesperson gain confidence. This study tries to understand the influence of listening to customer needs on their repurchase intentions. The study also tries to examine the role of persuasion skills of salespeople on customer. The study takes into account four interpersonal skills of the salesperson, including price delegation and stewardship and tries to examine the role of all four in developing interpersonal skills in salespeople further leading to influencing the purchase decision.

Hypothesis H1 tries to understand the influence of a salesperson's interpersonal skills on a customer's repurchase behavior. The study tries to explore if there exists a positive influence of paying attention to customer needs on product repurchase. Paying attention can generate better relationships leading to repurchase (Comer & Drollinger, 1999). Price plays an important in product purchase (Dixon & Schertzer, 2005). Customers get heavily influenced by-product pricing and pricing sometimes determine the firm's position (Dubinsky & Ingram, 1984). Price delegation is an important interpersonal skill which helps a salesperson to navigate through price negotiation.

It is important that a customer should be satisfied with the price he pays to the firm. Employee service plays an important role in customer satisfaction (Low et al., 2001). Paying attention to the customer and serving the customer appropriately can lead to multiple purchases of price sensitive products (McFarland, 2003). Do customers prefer Frontline employees to understand and interpret their needs and how often they feel that their needs are satisfied with the interpersonal skills of Frontline employees?

*3.3 Retail Operations*

**H2:** Retail operations negatively moderate the relationship between salesperson's interpersonal skills and customer repurchase intention.

Retail operations have been pivotal in developing customer experiences over time. The efficiency and effectiveness of a salesperson's action are two major factors that can shape a customer's purchase decision. Flawed retail operations manifest into lowered productivity, poor customer service, decreased employee engagement. Slow employees become a bottleneck during peak purchase periods, where long wait times are inevitable. Few Frontline employees result in a long wait time for consumers to complete their orders. It is very important to understand how listening abilities can suppress the negative effects of long wait times (Brown & Peterson, 1993). Long wait time, usually results in customer dissatisfaction further resulting in abandoning repurchases (Castleberry & Shepherd, 1993).

The list of available sets of products has influenced the purchase intention of customers for a long period (Castleberry & Shepherd, 1993). A long list of product range helps customers to make better choices in their selection and don't find themselves being pushed to purchase products (Comer & Drollinger, 1999). This leads to multiple repurchases further leading to loyalty towards a brand or the retailer. Having a long-range of products influences the purchasing ability of the customer (Ramsey & Sohi, 1997). This study tries to find out the extent of influence long product range has on customer repurchase decision. Overall, this study tries to measure the influence of the product range as an observed variable for the construct retail operations on customer repurchase intention.

It is very important to understand how salesperson's interpersonal skills can lead to better customer satisfaction (Chonko & Burnett, 1983). The hypothesis H2 tries to examine how retail operations play a moderating role on salesperson's interpersonal skills and customer repurchase intention. Hypothesis H2 tries to understand the influence the moderator retail operations have on customer repurchase intention. The study tries to explore the negative influence of retail operations on salesperson's interpersonal skills and customer repurchase intention. The hypothesis aims to discover if the customers are more or less willing to repurchase products if flawed retail operations create constraints during their purchase journey.

*3.4 Hard-Sell Sales Techniques*

**H3:** Hard-sell sales techniques negatively moderate the relationship between a salesperson's interpersonal skills and customer repurchase intention.

Many retailers in the past have been aggressive in their sales approach using sales performance as a benchmarking tool for salesperson's appraisal. Hard-selling strategies have been the common approach that many retailers took in the past two decades to increase their revenues. What constitutes hard-sell is the nature of the salesperson and his appeal in making the sales call. Many salespersons in spite of having been trained on sales techniques impose their views on customers in form of redundant information. Such information leads to dissatisfaction among customers further leading to customer abandonment. Hard-sell sales are an important aspect of a sales strategy that can act as a double-edged sword. It can lead to higher sales, at the same time can turn away customers due to its pushy nature.

It is important to understand the role of the hard-sell sales technique on the relationship between a salesperson's interpersonal skills and customer repurchase intention. We in our study try to answer this question by examining the moderating role of hard-sell sales techniques. Hypothesis H3 tries to examine the moderating role of hard-sell sales techniques on the relationship between salesperson's interpersonal skills and customer repurchase intentions. We believe that the moderating effect would have a negative effect on the relationship. The hypothesis is aimed to discover this negative moderating effect.

## 4. Data collection

The data collection was done using survey methodology. A sample consisting of a panel of consumers who interact with Frontline employees for purchase was selected from the responses to the survey screener. A quantitative questionnaire was designed in this phase. The questionnaire was divided into two parts, it contained a screener followed by the main section. The survey was administered using an online survey instrument. Data was analyzed using SPSS. Overall quantitative methods were used to analyze the data.

Table 1. Variable Descriptions

| Variable | Description |
| --- | --- |
| Active Listening | Continuous variable using interval scale examining the listening skills of salespeople from customer's perspective. |
| Persuasion | Continuous variable using interval scale examining the salesperson's persuasion from customer's perspective. |
| Delegation | Continuous variable using interval scale examining the salesperson's price delegation from customer's perspective. |
| Stewardship | Continuous variable using interval scale examining the salesperson's stewardship skills from customer's perspective. |
| Wait Times for Customer | Number of minutes a customer has to wait. |
| Slower Employees | Continuous variable using interval scale examining the salesperson's reflective skills from customer's perspective. |
| Lesser Product Range | Number of products offered in a particular category. |
| Disorganized Stores | Continuous variable using interval scale examining the organized store environment from customer's perspective. |
| Pushy Salespeople | Continuous variable using interval scale examining the salesperson's persistence towards purchase from customer's perspective. |
| Overwhelming Information | Continuous variable using interval scale examining the redundant information provided by salesperson to customers. |
| Lack of Emotional Appeal | Continuous variable using interval scale examining the lack of emotional appeal by salesperson to customers. |
| Customer Repurchase Intention | Continuous variable using interval scale measuring the likelihood of customer repurchase from the same store. |

The survey consisted of 26 questions (items) and took an average of 20 minutes to complete. Customers from five different retail outlets who belonged to the same company were used for this study. A screener was placed before the main survey to filter the respondents, who did visit the store but didn't purchase. The basic filter criteria were to select respondents who at least made a single purchase at the retail outlet. The selected respondents were sent an email link to the survey. The survey was distributed among 4500 respondents and 2332 completed responses were collected with a response rate of 51%. Only completed surveys were used for data analysis. Data was collected using online survey programmed in Qualtrics XM. The survey was distributed among a panel of retail shoppers. The panel respondent information was shared by the retail client who sponsored this study. The scales for the observed variable items were adapted from scales found in existing research that focuses on salesperson's interpersonal skills and customer repurchase intention.

## 5. Demographics and Exploratory Analysis

Table 2. Summary Statistics

| Variable | Mean | Std. dev | Min | Max |
| --- | --- | --- | --- | --- |
| Active Listening | 3.96 | 0.92 | 1 | 5 |
| Persuasion | 3.62 | 1.1 | 1 | 5 |
| Delegation | 3.51 | 1.14 | 1 | 5 |
| Stewardship | 3.38 | 1.12 | 1 | 5 |
| Wait times for customer | 6.07 | 4.02 | 0 | 16 |
| Slower Employees | 2.62 | 1.31 | 1 | 5 |
| Lesser Product Range | 10.05 | 5.28 | 1 | 22 |
| Pushy Salespeople | 3.88 | 1.3 | 1 | 5 |
| Overwhelming Information | 2.84 | 0.98 | 1 | 5 |
| Lack of Emotional Appeal | 3.68 | 1.62 | 1 | 5 |
| Customer Repurchase | 3.89 | 0.96 | 1 | 5 |

The survey consisted of a section that recorded demographic information of the respondents such as age, gender, marital status of the customer. The demographics recorded report that 66% were females and 34% were male. 40% of the respondent were full-time employed, 22% were part-time employed, 6% were unemployed, 20% were students and not employed, 12% percent were retired. 20% of respondents were in the age group of 18 to 25 years, 42% were in the age group of 26 to 35 years, 22% were in the age group of 36 to 50 years, and 16% were 51 years and above.

## 6. Research Methodology

*6.1 Multiple Linear Regression*

Equation for direct effect of Salesperson's interpersonal skills on Customer repurchase intention

$$Y_{Customer\ Repurchase\ Intention} = b_0 + b_1 x_{Salesperson's\ interpersonal\ skills} + \varepsilon_i$$

Equation for moderating effect of Retail operations on the direct effect

$$Y_{Customer\ Repurchase\ Intention} = b_0 + b_1 x_{Salesperson's\ interpersonal\ skills} + b_2 x_{Retail\ Operations} + b_3 x_{Salesperson's\ interpersonal\ skills} * x_{Retail\ Operations} + \varepsilon_i$$

Equation for moderating effect of Hard sell sales techniques on the direct effect

$$Y_{Customer\ Repurchase\ Intention} = b_0 + b_1 x_{Salesperson's\ interpersonal\ skills} + b_2 x_{Hard\ Sell\ Sales\ Techniques} + b_3 x_{Salesperson's\ interpersonal\ skills} * x_{Hard\ Sell\ Sales\ Techniques} + \varepsilon_i$$

Martins Gonçalves and Sampaio (2012) in their study on customer repurchase intention have examined the moderating effects of customer characteristics using multiple linear regression as a method to examine the effect. In another study by Fang et.al (2016), they examined the role of age and gender as a moderator on the relationship between perceived value and repurchase intention. In their study, they used multiple linear regression to examine the moderating effect. We have explored and examined multiple works of literature published in the last decade where they have successfully used multiple linear regression as an effective and efficient method for moderator analysis. Therefore, we concluded that in our research we are going to examine the moderation effect using multiple linear regression analysis.

Data were analyzed in two steps, the first being exploratory factor analysis using the principal component analysis method using varimax rotation to test for internal consistency. The second was moderation analysis using multiple linear regression analysis using the least squares method estimation. The analysis was done using SPSS Process as a statistical package. Exploratory factor analysis was used to test the homogeneity of each of the constructs using internal consistency analysis. For measuring the internal consistency of each construct Cronbach's alpha was used with a minimum alpha of 0.8.

For the second step causal research design was used to analyze the data. Causal research was also used to test the hypothesis. The cause-and-effect relationship was determined by analyzing the independent variable to check its effect on the dependent variable. Multiple linear regression analysis was used to examine the causal relationship between the dependent and independent variables and the moderation effect. Multiple linear regression was used for the methodological approach because it allows the dependent variable to be continuous and predicts the outcome with a moderation effect.

## 7. Results

Table 3. Exploratory Factor Analysis

| Factor | Variables | Factor Loading |
|---|---|---|
| 1 | **Salesperson's interpersonal skills** | |
| | Active Listening | 0.92 |
| | Persuasion | 0.78 |
| | Delegation | 0.62 |
| | Stewardship | 0.64 |
| | ***Cronbach's Alpha (0.88)*** | |
| 2 | **Retail operations** | |
| | Wait times for customer | 0.86 |
| | Slower Employees | 0.58 |
| | Lesser Product Range | 0.62 |
| | Disorganized Stores | 0.66 |
| | ***Cronbach's Alpha (0.86)*** | |
| 3 | **Hard-sell sales technique** | |
| | Pushy salespeople | 0.72 |
| | Overwhelming information | 0.56 |
| | Lack of emotional appeal | 0.68 |
| | ***Cronbach's Alpha (0.84)*** | |

Note: Principal component analysis method using Varimax rotation

We in this study have used exploratory factor analysis using the principal component analysis method and varimax rotation as step 1 to encounter the challenge that arises from issues in internal consistency. To test the homogeneity of each of the constructs for internal consistency analysis we used exploratory factor analysis. For measuring the internal consistency of each construct Cronbach's alpha was used with a minimum alpha of 0.8. We conclude that all the observed variables load on their respective constructs with Cronbach's alpha value higher than 0.8. As per our research framework and theoretical foundation, we derived three factors named as salesperson's interpersonal skills, retail operations, and hard-sell sales technique. The derived constructs were used in linear regression analysis to test the direct and moderator effect.

Table 4. Simple Linear Regression

Model Summary

| Model | R | R Square | Adjusted R Square | Std. Error | F | Sig |
|---|---|---|---|---|---|---|
| 1 | 0.926 | 0.857 | 0.838 | 4.32 | 51.7 | .000 |

a. Predictors: (Constant), Salesperson's interpersonal skills (SIS)

b. Dependent variable: Customer repurchase intention

| | Model | Unstandardized Coefficients | | Standardized Coefficients | | |
|---|---|---|---|---|---|---|
| | | B | Std. Error | Beta | t | Sig |
| 1 | (constant) | 8.72 | 0.168 | | 11.64 | .000 |
| | Salesperson's interpersonal skills (SIS) | 3.86 | 0.778 | 1.72 | 7.33 | .000 |

Note: Variables are tested at 1% (0.01) significance level; dependent variable: Customer repurchase intention

To measure the direct effect, we have used simple linear regression to examine if a salesperson's interpersonal skills (SIS) positively influence customer repurchase intention. From the analysis in table 4, we conclude that there exists a significant positive relationship between them. The r-square value 0.857 at a 1% significance level indicates that the positive relationship is highly significant. Thus, we conclude that our hypothesis 1 stands to its claim and cannot be rejected.

Table 5. Multiple Linear Regression Analysis and Moderator: Retail Operations (RO)

Model Summary

| Model | R | R Square | Adjusted R Square | Std. Error of the Estimate | Change Statistics | | |
|---|---|---|---|---|---|---|---|
| | | | | | R square change | F change | Sig. F Change |
| 1 | 0.927[a] | 0.886 | 0.874 | 5.356 | 0.886 | 60.76 | .000 |
| 2 | 0.966[b] | 0.946 | 0.927 | 3.316 | 0.064 | 21.07 | .000 |

a. Predictors: (Constant), Salesperson's interpersonal skills (SIS), Retail operations (RO)

b. Predictors: (Constant), Salesperson's interpersonal skills (SIS), Retail operations (RO), SIS x RO

| Model | | Unstandardized Coefficients | | Standardized Coefficients | | |
|---|---|---|---|---|---|---|
| | | B | Std. Error | Beta | t | Sig |
| 1 | (constant) | 21.35 | 0.486 | | 9.28 | .000 |
| | Salesperson's interpersonal skills (SIS) | 2.13 | 0.261 | 0.61 | 11.63 | .000 |
| | Retail operations (RO) | -18.64 | 0.166 | -1.38 | -7.44 | .000 |
| 2 | (constant) | 32.28 | 0.238 | | 8.09 | .000 |
| | Salesperson's interpersonal skills (SIS) | 0.96 | 0.068 | 0.82 | 6.72 | .223 |
| | Retail operations (RO) | -16.11 | 0.071 | -0.38 | -11.88 | .000 |
| | SIS x RO | -3.036 | 0.008 | -0.29 | -8.52 | .000 |

Note: Variables are tested at 1% (0.01) significance level; dependent variable: Customer repurchase intention

Table 6. Coefficients for Moderator Retail Operation's Observed variables

| Retail Operations Group | Model | Unstandardized Coefficients | | Standardized Coefficients | t | Sig | Correlations | | |
|---|---|---|---|---|---|---|---|---|---|
| | | B | Std. Error | Beta | | | Zero order | Partial | Part |
| Wait times for customers | (constant) | 34.27 | 0.668 | | 52.1 | 0.00 | 0.558 | 0.558 | 0.558 |
| | Salesperson's interpersonal skills | 1.86 | 0.256 | 0.768 | 7.65 | 0.00 | | | |
| Slower employees | (constant) | 32.62 | 0.873 | | 32.6 | 0.00 | 0.784 | 0.784 | 0.784 |
| | Salesperson's interpersonal skills | 1.38 | 0.252 | 0.589 | 4.96 | 0.00 | | | |
| Lesser product range | (constant) | 37.56 | 0.856 | | 64.4 | 0.00 | 0.082 | 0.082 | 0.082 |
| | Salesperson's interpersonal skills | 0.75 | 0.376 | 0.684 | 4.92 | 0.45 | | | |
| Disorganized stores | (constant) | 36.72 | 0.724 | | 57.4 | 0.00 | 0.388 | 0.388 | 0.388 |
| | Salesperson's interpersonal skills | 1.16 | 0.158 | 0.328 | 1.31 | 0.00 | | | |

Note: Variables are tested at 1% (0.01) significance level; dependent variable: Customer repurchase intention

In this study, we relied on multiple linear regression for moderation analysis using retail operations as our first moderator. We examined if retail operations (RO) negatively influence the positive relationship between salesperson's interpersonal skills (SIS) and customer repurchase intention. From the analysis in table 5, we conclude that there exists a significant negative moderation effect of retail operations on the positive relationship between salesperson's interpersonal skills and customer repurchase intention. The change in r-square value 0.064 at a 1% significance level indicates that the negative moderation effect is highly significant. Thus, we conclude that our hypothesis 2 stands to its claim and cannot be rejected.

We also analyze the moderating effects of individual observed variables of the construct retail operations on the relationship (from table 6). We conclude that slower employees and long wait times for customers are the significant observed variables having a stronger moderating effect as retail operations on the positive direct relationship between salesperson's interpersonal skills (SIS) and customer repurchase intention. Disorganized stores were found to be significant but its effect is negligible.

Table 7. Multiple Linear Regression Analysis and Moderator: Hard-Sell Sales Techniques (HSST)

Model Summary

| Model | R | R Square | Adjusted R Square | Std. Error of the Estimate | Change Statistics | | |
|---|---|---|---|---|---|---|---|
| | | | | | R square change | F change | Sig. F Change |
| 1 | 0.877 | 0.736 | 0.774 | 6.048 | 0.658 | 22.82 | .000 |
| 2 | 0.906 | 0.847 | 0.857 | 4.161 | 0.182 | 14.71 | .000 |

a. Predictors: (Constant), Salesperson's interpersonal skills (SIS), Hard-Sell sales technique (HSST)

b. Predictors: (Constant), Salesperson's interpersonal skills (SIS), Hard-Sell sales technique (HSST), SIS x HSST

| | Model | Unstandardized Coefficients | | Standardized Coefficients | | |
|---|---|---|---|---|---|---|
| | | B | Std. Error | Beta | t | Sig |
| 1 | (constant) | 33.56 | 1.658 | | 8.64 | .000 |
| | Salesperson's interpersonal skills (SIS) | 3.46 | 0.926 | 0.78 | 11.23 | .000 |
| | Hard-Sell sales technique (HSST) | -28.61 | 1.238 | -0.52 | -8.64 | .000 |
| 2 | (constant) | 23.58 | 0.335 | | 7.81 | .000 |
| | Salesperson's interpersonal skills (SIS) | 1.68 | 0.178 | 0.54 | 8.23 | .386 |
| | Hard-Sell sales technique (HSST) | -22.41 | 0.831 | -0.67 | -11.53 | .000 |
| | SIS x HSST | -11.61 | 0.782 | -0.83 | -14.52 | .000 |

Note: Variables are tested at 1% (0.01) significance level; dependent variable: Customer repurchase intention

Table 8. Coefficients for Moderator Hard-Sell Sales Technique (HSST) Observed variables

| Hard-Sell Sales Operation Group | Model | Unstandardized Coefficients | | Standardized Coefficients | t | Sig | Correlations | | |
|---|---|---|---|---|---|---|---|---|---|
| | | B | Std. Error | Beta | | | Zero order | Partial | Part |
| Pushy salespeople | (constant) | 24.37 | 0.527 | | 42.3 | 0.00 | 0.734 | 0.734 | 0.734 |
| | Salesperson's interpersonal skills | 4.16 | 0.058 | 0.838 | 5.73 | 0.00 | | | |
| Overwhelming information | (constant) | 22.42 | 0.632 | | 22.3 | 0.00 | 0.627 | 0.627 | 0.627 |
| | Salesperson's interpersonal skills | 2.38 | 0.185 | 0.642 | 7.92 | 0.00 | | | |
| Lack of emotional appeal | (constant) | 27.31 | 0.752 | | 44.6 | 0.00 | 0.438 | 0.438 | 0.438 |
| | Salesperson's interpersonal skills | 3.23 | 0.273 | 0.583 | 6.52 | 0.00 | | | |

Note: Variables are tested at 1% (0.01) significance level; dependent variable: Customer repurchase intention

For the second moderator hard-sell sales technique, we followed the same analysis that we conducted for the first moderator. We examined if hard-sell sales techniques (HSST) negatively influence the positive relationship between salesperson's interpersonal skills (SIS) positively and customer repurchase intention. From the analysis in table 7, we conclude that there exists a significant negative moderation effect of hard-sell sales techniques on the positive relationship between salesperson's interpersonal skills and customer repurchase intention. The change in r-square value 0.182 at a 1% significance level indicates that the negative moderation effect is highly significant. Thus, we conclude that our hypothesis 3 stands to its claim and cannot be rejected.

We also analyze the moderating effects of individual observed variables of the construct hard-sell sales techniques on the relationship (from table 8). We conclude that pushy salespeople and overwhelming information are the significant observed variables having a stronger moderating effect as hard-sell sales techniques on the positive main effect. Lack of emotional appeal is found to be significant but its effect is negligible.

## 8. Discussion

Table 9. Result of the Hypothesis Testing

| No. | Hypothesis | Results |
|---|---|---|
| 1 | **H1:** Salesperson's interpersonal skills positively influence customer repurchase intention. | Supported |
| 2 | **H2:** Retail operations negatively moderate the relationship between salesperson's interpersonal skills and customer repurchase intention. | Supported |
| 3 | **H3:** Hard-sell sales techniques negatively moderate the relationship between a salesperson's interpersonal skills and customer repurchase intention. | Supported |

This study has several contributions to the literature of salesperson's interpersonal skills in the field of personal selling and sales management. We have introduced two major moderators, retail operations and hard-selling strategies in this study. This study contributes to the existing literature by examining the significant negative moderating effects of retail operations and hard-selling strategies over salesperson's interpersonal skills and customer repurchase intention. The theory of service management has been applied to retail management in this study to examine the tangible effects of operations on customer's repurchase behavior. The application of this theory in this study helps open up new streams of research in retail sales management.

It is inferred from this study that slower employees and longer wait times contribute heavily to flawed retail operations making the process inefficient and further hindering the salesperson's performance. It is also being examined in this study that pushy salespeople and hard-selling techniques lead to failed sales strategies. It is of the utmost importance that the retail industry today should accommodate extraneous factors when developing better corporate strategies.

## 9. Industrial Implications

This study helps managers to understand the influence of retail operations and sales strategy on sales performance. Until now it was believed that salesperson's interpersonal skills are the sole contributors to sales performance, which invariantly lead to rise and decline in sales. This study has important findings for the retail industry. Since we get significant negative moderating effects from retail operations and hard-selling strategies, this conclusion contradicts the traditional belief. For retail managers, this study can help them develop better sales strategies thinking beyond hard-selling, also for them, this can help build better retail operations.

## 10. Conclusion

This study helps us to understand how a retail organization builds strategies along their Frontline employees to improve their sales through customer repurchase. This study proves that retail operations and hard-sell sales techniques both have a negative influence on the relationship between a salesperson's interpersonal skills and customer repurchase intention. Salesperson's interpersonal skills can motivate customers to repurchase products from the same retail outlet, but retail operations can be pragmatic constraints to the process. Hard-sell sales techniques also hinder the repurchase decision by creating an emotional gap between customers and salespeople. The customer won't feel being pushed over to make choices over a confined range due to the pushy nature of the salespeople. This study adds to the existing literature by filling the research gaps by introducing retail operations and hard-sell sales techniques into the current relationship. This helps us to better understand the influence of retail operations and hard-sell sales techniques over customer repurchase decisions.

Eventually, the study helps us to understand that slower employees can create long wait times further negatively influencing a customer's repurchase decision. Training employees to increase efficiency by focusing on the needs of the customers and effectively reducing the wait time can help retain customers. Organizations highly depend upon salesperson-consumer interaction for sales performances. This study will help firms to improve their employee training and make their organizations more customer-oriented. The results indicate if organizations can allocate a sufficient amount of budget to train Frontline sales personal, they can improvise on skill improvements of the employees. Listening to the needs of the customer is one of the strongest skills set that can lead to a better customer relationship, further leading to higher chances of repurchase.

This experience can add up to developing higher loyalty in customers and improves their repurchase behavior. This can also make the Frontline employees more efficient in processing customer needs and improvise the store's operational and logistical efficiencies. This study also helps organizations to understand the relevance of retail operations, in the context of repurchase behavior and increased customer loyalty. Promotional efforts can be focused on certain groups who exhibit a higher tendency towards repurchase thus further influencing the loyalty but by soft-sell rather than hard-sell sales techniques. Overall, this study helps organizations to better allocate budget and efforts towards effective sales strategies.

**References**

Anderson, J. C., Narus, J. A., & Narayandas, D. (2008). *Business market management: Understanding, creating, and delivering value,* 3rd ed. Upper Saddle River (NJ): Prentice Hall.

Babakus, Emin, David W. Cravens, Mark Johnston, and William C. Moncrief (1999). *The Role of Emotional Exhaustion in Sales Force Attitude and Behavior Relationships. Journal of the Academy of Marketing Science, 27*(1), 58–70. https://doi.org/10.1177/0092070399271005

Bello, K.B., Jusoh, A., & Md Nor, K. (2020). Relationships and impacts of perceived CSR, service quality, customer satisfaction and consumer rights awareness. *Social Responsibility Journal,* Vol. ahead-of-print No. ahead-of-print. https://doi.org/10.1108/SRJ-01-2020-0010

Ben Amor, N. E. (2019). What skills make a salesperson effective? An exploratory comparative study among car sales professionals. *International Business Research, 12*(11), 76. https://doi.org/10.5539/ibr.v12n11p76

Bhuian, Shahid N., Bulent Menguc, & Rene Borsboom. (2005). Stressors and Job Outcomes in Sales: A Triphasic Model Versus a Linear-Quadratic-Interactive Model. *Journal of Business Research, 58*(2) (February), 141–150. https://doi.org/10.1016/S0148-2963(03)00132-2

Boles, James S., Mark W. Johnston, and Joseph F. Hair, Jr. (1997). Role Stress, Work–Family Conflict and Emotional Exhaustion: Inter-Relationships and Effects on Some Work-Related Consequences. *Journal of Personal Selling & Sales Management, 17*(1) (Winter), 17–28.

Brown, Steven P., and Robert A. Peterson. (1993). Antecedents and Consequences of Salesperson Job Satisfaction: Meta- Analysis and Assessment of Causal Effects. *Journal of Marketing Research, 30* (February), 63–77. https://doi.org/10.1177/002224379303000106

Butt, I., Mukerji, B., & Shareef, M.A. (2017). Relevance of soft-sell and hard-sell advertising appeals for global consumer cultural positioning. *Journal of Customer Behaviour, 16*(3), 263-279. https://doi.org/10.1362/147539217X15071081721125

Castleberry, Stephen B., & C. David Shepherd. (1993). Effective Interpersonal Listening and Personal Selling. *Journal of Personal Selling and Sales Management, 18*(1), 35-49.

Chonko, Lawrence B., & John Burnett. (1983). Measuring the Importance of Ethical Situations as a Source of Role Conflict: A Survey of Salespeople, Sales Managers, and Sales Support Personnel. *Journal of Personal Selling & Sales Management, 3*(1) (May), 41–47.

Churchill, Gilbert A., Jr., Neil M. Ford, Steven W. Hartley, & Orville C. Walker, Jr. (1985). The Determinants of Salesperson Performance: A Meta-Analysis. *Journal of Marketing Research, 22*(May), 103–118. https://doi.org/10.1177/002224378502200201

Comer, Lucette B., & Tanya Drollinger (1999). Active Empathetic Listening and Selling Success: A Conceptual Framework. *Journal of Personal Selling and Sales Management, 19*(1), 15-29.

Curina, I., Francioni, B., Hegner, S. M., & Cioppi, M. (2020). Brand hate and non-repurchase intention: A service context perspective in a cross-channel setting. *Journal of Retailing and Consumer Services, 54,* 102031. https://doi.org/10.1016/j.jretconser.2019.102031

Dick, A.S. and Basu, K. (1994). Customer Loyalty: Toward an Integrated Conceptual Framework. *Journal of the Academy of Marketing Science, 22*(2), 99–113. https://doi.org/10.1177/0092070394222001

Dixon, Andrea L., and Susan M.B. Schertzer (2005). Bouncing Back: How Salesperson Optimism and Self-Efficacy Influence Attributions and Behaviors Following Failure. *Journal of Personal Selling & Sales Management, 25*(1), 361–369.

Du, H., Xu, J., Tang, H., & Jiang, R. (2020). Repurchase Intention in Online Knowledge Service: The Brand Awareness Perspective. *Journal of Computer Information Systems,* In-Print. https://doi.org/10.1080/08874417.2020.1759159

Dubinsky, Alan J., & Thomas N. Ingram (1984). Correlates of Salespeople's Ethical Conflict: An Exploratory Investigation. *Journal of Business Ethics, 3* (November), 343–353. https://doi.org/10.1007/BF00381759

Eisenhardt, K. M. (1989). Agency Theory: An Assessment and Review. *The Academy of Management Review, 14*(1), 57. https://doi.org/10.5465/amr.1989.4279003

Fullerton, R. A. (1988). How modern is modern marketing? Marketing's evolution and the myth of the production era. *Journal of Marketing, 52*(1), 108–125. https://doi.org/10.1177/002224298805200109

Funkhouser, G. R. (1984). A Practical Theory of Persuasion Based on Behavioral Science Approaches. *Journal of Personal Selling & Sales Management, 4*:2, 17-25.

Gabler, C. B., Vieira, V. A., Senra, K. B., & Agnihotri, R. (2019). Measuring and testing the impact of interpersonal mentalizing skills on retail sales performance. *Journal of Personal Selling & Sales Management, 39*(3), 222-237. https://doi.org/10.1080/08853134.2019.1578661

Goad, Emily A. (2014). *The Impact of Salesperson Listening: A Multi-faceted Research Approach (Doctoral dissertation).* The University of Texas at Arlington, Texas, USA.

Handayani, P., Ariantana, E., & Pinem, A. (2020). How to Increase Customer Repurchase Intention in a Business-to-Customer (B2C)? *International Journal of Electronic Commerce Studies, 11*(1), 1- 19. https://doi.org/10.7903/ijecs.1721

Ho, M. H., & Chung, H. F. (2020). Customer engagement, customer equity and repurchase intention in mobile apps. *Journal of Business Research, 121,* 13-21. https://doi.org/10.1016/j.jbusres.2020.07.046

Ilyas, GB., Rahmi, S., Tamsah, H., Munir, AR, & Putra, AHPK (2020). Reflective Model of Brand Awareness on Repurchase Intention and Customer Satisfaction. *The Journal of Asian Finance, Economics and Business, 7*(9), 427–438. https://doi.org/10.13106/jafeb.2020.vol7.no9.427

Ingram, Thomas N., Schwepker, Jr., Charles H., & Don Hutson (1992). Why Salespeople Fail. *Industrial Marketing Management, 21,* 225-230. https://doi.org/10.1016/0019-8501(92)90019-P

Javed, M. K., & Wu, M. (2020). Effects of online retailer after delivery services on repurchase intention: An empirical analysis of customers' past experience and future confidence with the retailer. *Journal of Retailing and Consumer Services, 54,* 101942. https://doi.org/10.1016/j.jretconser.2019.101942

Kirca, A. H., Jayachandran, S., & Bearden, W. O. (2005). Market orientation: A Metanalytic Review and Assessment of its Antecedents and Impact on Performance. *Journal of Marketing, 69*(2), 24–41. https://doi.org/10.1509/jmkg.69.2.24.60761

Liu, Y., Hochstein, B., Bolander, W., Bradford, K., & Weitz, B. A. (2020). Internal selling: Antecedents and the importance of networking ability in converting internal selling behavior into salesperson performance. *Journal of Business Research, 117,* 176-188. https://doi.org/10.1016/j.jbusres.2020.04.036

Low, George S., David W. Cravens, Ken Grant, & William C. Moncrief (2001). Antecedents and Consequences of Salesperson Burnout. *European Journal of Marketing, 35*(5–6), 587–611. https://doi.org/10.1108/03090560110388123

Martins Gonçalves, H., & Sampaio, P. (2012). The customer satisfaction-customer loyalty relationship. *Management Decision, 50*(9), 1509–1526. https://doi.org/10.1108/00251741211266660

McFarland, Richard G. (2003). Crisis of Conscience: The Use of Coercive Sales Tactics and Resultant Felt Stress in the Salesperson. *Journal of Personal Selling & Sales Management, 23*(4) (Fall), 311–325.

Mowen, John C., Minor, Michael. (1998). *Consumer Behavior E5* (5th ed). New York: PRENTICE-HALL.

Munnukka, J. (2008). Customers' Purchase Intentions as a Reflection of Price Perception. *Journal of Product & Brand Management, 17*(3), 188-196. https://doi.org/10.1108/10610420810875106

Naumann, Earl, Scott M. Widmier, and Donald W. Jackson (2000). *Examining the Relationship Between Work Attitudes and Propensity to Leave Among Expatriate Salespeople.* Journal of Personal Selling & Sales Management, 20, 4 (Fall), 227–241.

Pan, Y., Sheng, S., & Xie, F. (2012). Antecedents of Customer Loyalty: An Empirical Synthesis and Reexamination. *Journal of Retailing and Consumer Services, 19*(1), 150-158. https://doi.org/10.1016/j.jretconser.2011.11.004

Pelham, Alfred M. (2009). An Exploratory Study of the Influence of Firm Market Orientation on Salesperson Adaptive Selling, Customer Orientation, Interpersonal Listening in Personal Selling and Salesperson Consulting Behaviors. *Journal of Strategic Marketing, 17*(1), 21-39. https://doi.org/10.1080/09652540802619202

Ramsey, Rosemary P., & Ravipreet S. Sohi (1997). Listening to Your Customers: The Impact of Perceived Salesperson Listening Behavior on Relationship Outcomes. *Journal of the Academy of Marketing Science, 25*(2), 127-137. https://doi.org/10.1007/BF02894348

Saxe, R., & Weitz, B. A. (1982). The SOCO Scale: A Measure of the Customer Orientation of Salespeople. *Journal of Marketing Research, 19*(3), 343-351. https://doi.org/10.1177/002224378201900307

Simanjuntak, M., Nur, H., Sartono, B & Sabri, M. (2020). A general structural equation model of the emotions and repurchase intention in modern retail. *Management Science Letters, 10*(4), 801-814. https://doi.org/10.5267/j.msl.2019.10.017

Singh, B., & Tiwari, A. A. (2019). Customer stewardship behavior and stewardship fatigue: a conceptual framework. *Marketing Intelligence & Planning, 38*(3), 386-399. https://doi.org/10.1108/MIP-02-2019-0071

Slack, N., Singh, G. & Sharma, S. (2020). *The effect of supermarket service quality dimensions and customer satisfaction on customer loyalty and disloyalty dimensions. International Journal of Quality and Service Sciences, 12*(3), 297-318. https://doi.org/10.1108/IJQSS-10-2019-0114

Tandon, U., Mittal, A. & Manohar, S. (2020). Examining the impact of intangible product features and e-commerce institutional mechanics on consumer trust and repurchase intention. Electron Markets. https://doi.org/10.1007/s12525-020-00436-1

Udayana, I. B., Ardyan, E., & Farida, N. (2019). *Selling relationship quality to increase salesperson performance in the pharmacy industry. International Journal of Services and Operations Management, 33*(2), 262. https://doi.org/10.1504/ijsom.2019.10022077